\documentclass{iopart}
\usepackage{graphicx}
\makeatletter
\renewcommand{\@makefnmark}{\mbox{$^{\@thefnmark}$}}
\renewcommand{\@makefntext}[1]{\noindent\makebox[0.5em][r]{\@makefnmark}\ \ #1}
\makeatother

\begin{document}
\title[Mutual transformation among  bound, virtual and resonance states]{Mutual transformation among  bound, virtual  and resonance states in one-dimensional rectangular potentials}
\author{M Kawasaki$^1$, T Maehara$^2$ and M Yonezawa$^3$}
\address{$^1$ Physics Department, Gifu University, Yanagido, Gifu 501-1193, Japan}
\address{$^2$ Graduate School  of Education, Hiroshima University, 
Higashi-Hiroshima 739-8524, Japan}
\address{$^3$ Nakano 7-5-28, Aki-ku, Hiroshima 739-0321, Japan }
\ead{kawasaki@gifu-u.ac.jp, tmaehar@hiroshima-u.ac.jp and m-yonezawa@mtc.biglobe.ne.jp}
\date{\today}
%% ******************************************************************* % 
\begin{abstract}
A detailed analysis has been made by  R. Zavin and N. Moiseyev (2004 {\it J. Phys. A: Math, Gen,} \textbf{37} 4619) for the  change of  bound states into resonance states  via  coalescence of virtual  states in a one-dimensional symmetric rectangular attractive potential as it becomes shallow,   with convergent  wave functions of virtual and resonance states by the complex scaling method.  As a complement  to such an analysis,  we discuss some global features of the  pole spectrum  of  the $S$-matrix  by using a complex extension of the  real potential $V^{(\mathrm{real})}$  to $\rme^{\rmi \alpha}V^{(\mathrm{real})}$ with a real phase $\alpha$.  We show the structures of trajectories  of poles developed  for the change of $\alpha$ in the complex momentum plane, which is useful to understand the mutual transformation among the bound, virtual and resonance states.
 \end{abstract}
%% ******************************************************************* % 
 %\submitto{\JPA}
\maketitle
\setcounter{footnote}{0}
%
%% ******************************************************************* % 
\section{Introduction}
%% ******************************************************************* % 
In  the quantum mechanics the  dynamical properties of a system is determined by  its potential.    The bound states, virtual (antibound) states,  resonance states   with conjugate  complex-virtual (antiresonance) states produced by the potential  form  a basis   characterizing  the system. These states transform mutually as the potential  strength changes. 

Recently it has been discussed by  Zavin and Moiseyev  \cite{zvms}  how the bound states change  into  the virtual   states and subsequently into the resonance states as the potential depth  becomes shallow for a one-dimensional symmetric rectangular well potential.  The authors  studied  the problem about solutions of the Schr\"{o}dinger equation  with the complex scaling method;  this procedure gives  convergent wave functions  for the virtual and resonance states, which  otherwise  are divergent.

The poles of the $S$-matrix  in the complex  energy plane generated by the potential   correspond to the  bound, virtual, resonance  and complex virtual states.  An extensive study of the poles of the $S$-matrix element of a rectangular potential  was done previously by  Nussenzveig \cite{nssn}.  The main interest of these  analyses \cite{zvms,nssn} seems to be in the movements of various states or poles with the change of the potential depth.   As a complement to these analyses,   it will be useful to explore  a whole spectrum of the  poles for a given potential and to examine the change of spectrum with the potential depth.    The approach taken by Nussenzveig, however,  is very  intricate and  is not suitable for showing a  global picture  of the pole  spectrum. 

We examine this problem by a  systematic  way of  locating the resonance states as well as the bound and virtual states  for a given potential and show the change of the pole spectral  chart for  the one-dimensional rectangular potential.   This approach uses a complex extension of the real potential.  It  provides   a simple way of presenting an overall view  of the pole spectrum and is applicable to other potentials.  

There have been also many investigations which involve  complex extensions  of potentials, developed and widely used   in  calculations of resonance properties in various fields of  physics \cite{cxsmrl, cmxcd, riss, kkln}.   It seems, however, that  no attempt  has  been made  to use the complex extension for  exploring some global feature of the pole  spectrum as will be given in this article.

%% ******************************************************************* % 
\section{$S$-matrix }
%% ******************************************************************* % 
\subsection{Some relations of $S$-matrix elements}
The properties of $S$-matrix for a one-dimensional system with real potential  have been extensively studied \cite{nwtn}. There seems, however, practically   no article explicitly referring to  those of one-dimensional  system with complex potential to the knowledge of the present authors,  though  some detailed studies have been made  for the basic properties of poles of the $S$-matrix for the spherical symmetric  complex potential in three dimensional space \cite{cxsmrl}.   We, therefore, briefly derive  some properties of the $S$-matrix for the  one-dimensional complex potential.

 We  consider  a particle of mass $m$  in a complex potential $V(x)$  in one dimension. 
Here the potential  is  a finite-range potential given by
\begin{equation}
 V(x) =\gamma V^{(\mathrm{real})}(x)= \Biggl\{
\begin{array}{ccc}
           0                    &  x<- a  &             ({\rm   region \,\, I}) \,, \\
\gamma v(x)         &  -a \le x \le a   & ({\rm  region \,\, II})\, , \\
          0                      &  x>a     &             ({\rm  region \,\, III}) \, ,
\end{array}
\end{equation} 
where $v(x)$ is a real function of $x$ and $\gamma$ is a complex constant.
 The  time-independent  Schr\"{o}dinger equation  for the wave function  $\psi(x)$ of the particle  is  in units of $\hbar=1$ for a given energy $E=k^{2}/2m$
\begin{equation}\label{eq: scheq}
-\frac{1}{2m} \frac{\rmd^2}{\rmd x^2}\psi (x) + V(x) \psi(x) = \frac{k^{2}}{2m} \psi(x) \,,
\end{equation}
where  $k$ is the momentum of the particle in the exterior  interaction free regions I and III.

The wave function  $\psi_{1}(x, k, \gamma)$ for a unit incident beam along the positive direction of the  $x$ axis is   expressed  in regions I and  III as 
\begin{equation}\label{eq: exwf1}
\psi_{1}(x,k,\gamma) = \bigg\{
\begin{array}{lcc}
\rme^{+ \rmi kx} + S_{21}(k,\gamma)\,\rme^{-\rmi kx} &  x<- a & ({\rm  region \,\,I})\,, \\
 S_{11}(k,\gamma) \, \rme^{+\rmi kx}  &   x > a                          &  ({\rm  region\,\, III}) \, ,
\end{array}
\end{equation} 
and another wave function  $\psi_{2}(x, k, \gamma)$   along the negative direction is
\begin{equation} \label{eq: exwf2}
\psi_{2}(x,k,\gamma) = \bigg\{
\begin{array}{lcc}
S_{22}(k,\gamma) \, \rme^{-\rmi kx}  &  x<- a &({\rm  region\,\, I}) \, ,\\
 \rme^{- \rmi kx} + S_{12}(k,\gamma) \, \rme^{+\rmi kx} &  x >a  & ({\rm  region\,\, III}) \, .
\end{array}
\end{equation} 
A matrix with its elements $S_{ij}$ $(i,j=1,2)$ is the $S$-matrix of the present problem. 

The momentum $k$ is real so far and we have to make an analytic continuation of the $S$-matrix in the complex momentum plane.  We  prove the following relations about the $S$ matrix for the complex potential of finite range
\numparts
\begin{eqnarray}
 & S^{\rm t}(k,\gamma) \,S(-k,\gamma)  = S(k,\gamma) \,S^{\rm t}(-k,\gamma)  =1 \, , \label{eq: rel1} \\
 & S^{\dagger}(k,\gamma) \, S(k^{*}, \gamma^{*}) =1 \,, \label{eq: rel2} \\
 &  S^{*}(-k^{*},\gamma^{*})   =  S(k,\gamma)  \,. \label{eq: rel3}
\end{eqnarray}
\endnumparts

  For the real potential the information of the $S$-matrix in the first quadrant of the complex $k$-plane is enough to determine  the $S$-matrix in the entire complex plane.  This  is not enough for the complex potential as the relations (\ref{eq: rel2}) and (\ref{eq: rel3}) show.

\subsubsection{ Relation \rm{(\ref{eq: rel1})}}
 The Wronskian of  two linearly-independent solutions of the homogeneous second order ordinary linear differential equation with no first order term is constant.  Using this property of the Wronskian for solutions of (\ref{eq: scheq}), we derive  those properties of the $S$-matrix.

First we consider the Wronskian of   $\psi_{1}(x,k,\gamma)$ and $ \psi_{2}(x,k,\gamma)$
\begin{equation}
W[ \psi_{1}(x,k,\gamma),\psi_{2}(x,k,\gamma)] =
\biggl|
\begin{array}{cc}
\psi_{1}(x,k,\gamma) & \psi_{2}(x,k,\gamma) \\
\psi_{1}^{'}(x,k,\gamma) & \psi_{2}^{'}(x,k,\gamma) 
\end{array}
\biggr| \,.
\end{equation}
This gives
\begin{equation}
W[\psi_{1}(x,k,\gamma), \psi_{2}(x,k,\gamma)]_{x<-a} = -2\rmi k S_{22}(k,\gamma)
\end{equation}
and 
\begin{equation}
W[\psi_{1}(x,k,\gamma), \psi_{2}(x,k,\gamma)]_{x>a} = -2\rmi k S_{11} (k,\gamma)\, .
\end{equation}
Since these two values of the Wronskian are the same, we have a relation
\begin{equation}\label{eq: sma}
S_{11}(k,\gamma) =S_{22}(k,\gamma)\,.
\end{equation} 
            
The wave functions  $\psi_{1}(x,-k,\gamma)$ and $ \psi_{2}(x,-k,\gamma)$ are also solutions of  (\ref{eq: scheq}).  We calculate the Wronskians involving these solutions.  From $W[\psi_{1}(x,k,\gamma), \psi_{1}(x,-k,\gamma)] $ we have 
\begin{equation} \label{eq: smb}
S_{11} (k,\gamma) \, S_{11} (-k,\gamma) +  S_{21}(k,\gamma) \, S_{21}(-k,\gamma) =1\,, 
\end{equation} 
 from $W[\psi_{1}(x,k,\gamma),\psi_{2}(x,-k,\gamma)]$
\begin{equation}\label{eq: smc}
S_{11} (k,\gamma) \, S_{12} (-k,\gamma)+ S_{21}(k,\gamma) \, S_{22}(-k,\gamma) =0 \,,
\end{equation}   
and from $W[\psi_{2}(x,k,\gamma),\psi_{2}(x,-k,\gamma)]$
\begin{equation}\label{eq: smd}
S_{12} (k,\gamma)\, S_{12} (-k,\gamma)+ S_{22}(k,\gamma)\, S_{22}(-k,\gamma) =1\,.
\end{equation}   
Combining (\ref{eq: smb}), (\ref{eq: smc}) and (\ref{eq: smd}) gives
\begin{equation}
S^{\rm t} (k,\gamma) \, S(-k,\gamma) =1\,.
\end{equation}   
Further  (\ref{eq: sma}) implies
\begin{equation}
S(k,\gamma) \, S^{\rm t} (-k,\gamma) =1\,.
\end{equation}   
This completes  the relation (\ref{eq: rel1}).

\subsubsection{ Relations \rm{(\ref{eq: rel2})} \it{and}  {\rm (\ref{eq: rel3})}}
   First we derive the relation (\ref{eq: rel3}) and note on the relation (\ref{eq: rel2})  at the end.  

Let us take the complex conjugate of (\ref{eq: scheq}).  The wave function
$\psi_{1}^{*}(x, k, \gamma)$ satisfies the Schr\"{o}dinger equation (\ref{eq: scheq})  with  a potential $\gamma^{*}v(x)$ for energy $(k^{2})^{*}/2m$. The exterior part of this wave function is given by 
\begin{equation}
\psi_{1}^{*}(x,k,\gamma) = \bigg\{
\begin{array}{lc}
\rme^{- \rmi k^{*}x} + S_{21}^{*}(k,\gamma) \, \rme^{+\rmi k^{*}x}   &  x<- a \, ,  \\
 S_{11}^{*}(k,\gamma) \, \rme^{-\rmi k^{*}x}   &  x >a \,,
\end{array}
\end{equation} 
which  is written as
\begin{equation}
\psi_{1}^{*}(x,k,\gamma) = \bigg\{
\begin{array}{lc}
\rme^{+ \rmi (-k^{*})x} + S_{21}^{*}(k,\gamma) \, \rme^{-\rmi (- k^{*})x}   &  x<- a  \, ,\\
 S_{11}^{*}(k,\gamma) \, \rme^{+\rmi(- k^{*})x}   &  x >a  \;.
\end{array}
\end{equation} 
This  expression  is seen to be equivalent to the solution $\psi_{1}(x,-k^{*},\gamma^{*})$ given by
\begin{equation}
\psi_{1}(x,-k^{*},\gamma^{*}) = \bigg\{
\begin{array}{lc}
\rme^{+ \rmi (-k^{*})x} + S_{21}(-k^{*},\gamma^{*}) \, \rme^{-\rmi (- k^{*})x}   &  x<- a \, , \\
 S_{11}(-k^{*},\gamma^{*}) \, \rme^{+\rmi(- k^{*})x}  &  x >a \, .
\end{array}
\end{equation} 
Since such a solution is unique, we have relations
\begin{equation}
S_{21}^{*} (k,\gamma)=S_{21} (-k^{*},\gamma) \,\,\,\,{\rm and}\,\,\,\, S_{11}^{*}(k,\gamma) =S_{11}(-k^{*},\gamma^{*}) \, .
\end{equation}   

By the same  arguments for $\psi_{2}^{*}(k, \lambda)$ we have
\begin{equation}
S_{12}^{*} (k,\gamma)=S_{12} (-k^{*},\gamma) \,\,\,\, {\rm and}\,\,\,\, S_{22}^{*}(k,\gamma) =S_{22}(-k^{*},\gamma^{*})\,.
\end{equation}   
Hence we have the relation (\ref{eq: rel3}).

Finally we take the relation (\ref{eq: rel2}).
It is obvious that  this relation is obtainable from the relations (\ref{eq: rel1}) and (\ref{eq: rel3}).

\subsection{The $S$-matrix elements for a symmetric rectangular potential}

For the symmetric rectangular well,  the potential is given by $v(x)=-U$ where $U$ is a positive constant. The general solution $\psi_{\rm II}$ in the region II is given by 
\begin{equation}
\psi_{\rm II} = A\,\rme^{\rmi Kx} + B\,\rme^{-\rmi Kx}\,,
\end{equation}
where $K$ is $\sqrt{k^{2}+2m \gamma U}$.   The arbitrary constants $A$ and $B$ are determined by the  continuity conditions of the  wave functions and their space derivatives at $x=\pm a$ for each of the exterior wave functions (\ref{eq: exwf1}) and (\ref{eq: exwf2}). This  gives the  $S$-matrix as

\begin{eqnarray}\label{eq: sabcd}
S &=&\frac{2 \exp (-2\rmi ka)\,  kK}{2kK\cos2Ka-\rmi (k^2+K^2)\sin2Ka}  \nonumber \\ 
    & &  \times \left(
\begin{array}{cc}
1 & \displaystyle-\rmi \frac{k^2-K^2}{2kK}\,\sin2Ka \\ \displaystyle -\rmi \frac{k^2-K^2}{2kK}\,\sin2Ka & 1
\end{array}
\right) \,.
\end{eqnarray}

The poles of the $S$-matrix  elements come from the zeros of their  common denominator.   It is easy to observe the condition giving the zeros of the denominator is   just  the same as  that   obtained  by the eigenvalue equation for the bound-state energy.

Since the present potential is symmetric for the  space inversion $x \rightarrow -x$,  it is better to discuss the symmetric and antisymmetric  states  for the space inversion separately.  
The  exterior wave functions  $\psi_{\pm}(x, k, \gamma)$ of  the symmetric and antisymmetric states are given by 
\begin{equation}
\psi_{\pm}(x,k,\gamma) = \bigg\{
\begin{array}{lcc}
\rme^{+\rmi kx}   \pm S_{\pm}(k,\gamma)\,\rme^{-\rmi kx} &  x<- a & ({\rm  region \,\,I})\, , \\
 S_{\pm}(k,\gamma) \, \rme^{+\rmi kx} \pm\rme^{-\rmi kx} &   x > a  &  ({\rm  region\,\, III}) \, .
\end{array}
\end{equation}

The $S$-matrix $\hat{S }$ on the basis of  the symmetric  and antisymmetric states is given by
\begin{equation}\label{eq: smsa}
\hat{S} =
\left(
\begin{array}{cc}
S_{+} & 0 \\
0 & S_{-}
\end{array}
\right)
\end{equation}
with
\begin{eqnarray}
  S_{+}&=& \exp (-2\rmi ka)   \biggl[\frac{ k \cos Ka + \rmi K \sin Ka }{k \cos Ka -\rmi K\sin Ka }\biggr] \, , \label{eq: splus} \\ 
  S_{-}&= &\exp (-2\rmi ka)   \biggl[\frac{ K \cos Ka + \rmi k \sin Ka }{K \cos Ka -\rmi k \sin Ka }\biggr] \, . \label{eq: sminus} 
\end{eqnarray}
Here $\hat{S}$ is related  to $S$ of (\ref{eq: sabcd}) by a similarity transformation  $\hat{S}= U S U^{-1} = U S U^{\rm t} $ with $U$  given as 
\begin{equation}\label{eq:stu}
U =\frac{1}{\sqrt{2}}
\left(
\begin{array}{rr}
1 & 1 \\
-1 & 1
\end{array}
\right)\,.
\end{equation}

For  representing the pole structures,   the complex momentum $k$ plane gives a simpler picture than   the complex energy  $E$ plane which has cut structure, though the energy variable may be more familiar than the momentum variable at interface with experiments.  Here we  note on the cut structure of the $S$-matrix in the complex $E$ plane. There is a  cut  starting  from  the branch point  $E=0$ to $\infty$.  We denote the Riemann sheet given by  Im$\,k > 0$ as $ I$ and  the sheet for Im$\,k < 0$  as  $II$, which are called the physical and the unphysical  sheet, respectively.

In the complex momentum $k$ plane the bound state appears on the positive imaginary axis and the virtual state on the negative imaginary axis, while the resonance state emerges in the lower  half-plane with positive real part (the fourth quadrant of the complex plane)  and the   complex virtual  state (antiresonance state) with negative real part (the third quadrant).

There appear an infinite number of poles for finite range  potential  for the rectangular potential \cite{nwtn},   with an infinite number of resonance states  and  a finite number of bound and virtual states.

 Since the $S$-matrix elements  are given in  the explicit analytical form (\ref{eq: splus}) and (\ref{eq: sminus}) for the rectangular potential,  it is simple to find the  pole locations of the bound and virtual states for a given  well depth.  As they lie on the imaginary momentum axis, we can find zeros of the denominators in the expressions of the $S_{+}$ and  $S_{-}$   by numerical calculations  with or without  the aid of a standard graphical method which is  often very helpful.

 The resonance-state poles,  however, are less easy to be located, as they are scattered in the complex momentum plane.  One way is to trace   each of virtual states descending on the imaginary momentum axis and finally  changing  into a resonance pole by coalescence with  another virtual state  ascending  on the imaginary axis as the potential  becomes shallow, as done  by Zavin and Moiseyev \cite{zvms}.  This requires some work  to draw  a whole pole chart in the complex momentum plane for a \textit{given} value $U$,  since we need a tracing of one virtual state for each of resonances starting  from a deeper  well.  There is a simple way to draw the spectrum chart of resonance  poles  which is given in the following.

\section{Pole charts}

\subsection{Method for locating  poles }

\subsubsection{Procedures}   
Although we  study the one-dimensional problem in this paper, the present approach  is not restricted to it.

(i)  First we  search for poles on the imaginary axis for a given  real  potential depth ($\gamma=1$).  We do the same for the sign reversed potential, i.e, the repulsive rectangular barrier with  its height equivalent to the depth of the well ($\gamma=-1$). This can be easily  calculated numerically  if the $S$-matrix elements are given analytically.

(ii)  Next  step is to trace each of poles on the imaginary momentum axis  for  making a complex rotation of  the potential  $V(x)$ by taking
\begin{equation}\label{eq: phase}
               \gamma=  \rme^{\rmi \alpha} \,,
\end{equation}
with  a real phase parameter  $\alpha$ starting from $\alpha =0\,(\pi)$ for the poles of the well (barrier).  This is performed  numerically  by changing $\alpha$ with a small step. The well and the barrier are analytically  related  by the present complex  extension of the potential and  their pole spectra have close connections. 

These procedures produce  either a closed trajectory  or an open trajectory  of infinite length in the complex momentum plane for each of poles on the imaginary momentum axis. The    trajectories obtained in this way  present a simple picture for the global structures of the pole spectrum of the system.

 Relation (\ref{eq: rel3}) implies that each of the trajectories is mirror symmetric about the imaginary momentum axis or  has a mirror  partner.

If  the location of a resonance pole is found in some way, it is possible to draw a pole trajectory starting from this pole with the complex extension $\rme^{\rmi \alpha}$.  In general, it is much simpler to locate the virtual-state and bound-state poles  emerging on the imaginary momentum axis rather than the resonance-state poles in the complex  plane.  This is certainly one merit of the present approach, but the most interesting  point is the trajectory structures exposed by the phase rotation (\ref{eq: phase}).

There appear bound states only  for the well, while  resonance and virtual states appear for  both of  the well and  the barrier.  Hence,  we call the states appearing for the well  ``attractive" and those for the barrier ``repulsive"  for simplicity.

 \subsubsection{Some comments on the present approach}

Transformations with the change of phase $ \Delta \alpha =\pm \pi$ can be considered as a mapping  of the set of   all   the  attractive poles  to the set of repulsive poles, and  with $\Delta \alpha =\pm2\pi$   a self-mapping onto itself. 

For the self-mapping  transformation, there  appear  invariant subsets for which the mapping closes within  each of subsets.  A closed trajectory corresponds to a finite invariant subset which  consists of  a finite number of  attractive and repulsive  poles on the trajectory, while an open trajectory to an infinite invariant subset. Tracing a pole  of the  invariant subset  by continuously changing $\alpha$  implies repeating a self-mapping  for every change of the phase  by $\Delta \alpha=\pm 2\pi$, therefore, it  provides a systematic way of  scanning  all elements of the subset. The self-mapping   is essentially a cyclic permutation among the elements of a finite invariant  subset  and  a sequential transposition of the elements  for  an infinite invariant subset along its trajectory.

If the potential depth  varies,  it  occurs  degeneracy of  two virtual states  or of  a pair of resonance and  complex virtual states.  At this point the mapping is  indefinite and  the degeneracy induces the change of two virtual states into a pair of  resonance  and  complex virtual states, or in  reverse order.  This implies the loss of  one-to-one correspondence between the  two virtual-state  poles and   the pair of  resonance-state and  complex virtual-state poles  and causes  rearrangement between two  invariant  subsets.    It is added that the rearrangement between the invariant subsets happens  also when the corresponding trajectories mutually contact off the imaginary momentum axis.

 In the one-dimensional rectangular potential  the degeneracy of states occurs always on the negative imaginary momentum axis at the  fixed point $k_\mathrm{c}=-\rmi/a$, which is the only double zero point of the denominator  of the $S$-matrix elements, common  to $S_{+}$ and $S_{-}$. 

It could  happen  that some closed trajectories do not cross the imaginary momentum axis as the well depth varies  for the assumed strength of the  potential.   These trajectories  consist of only either resonance-state  poles or complex virtual-state poles.   In fact, such  trajectories appear  for the $d$-wave states in the spherical rectangular potential.  This, however, does not cause a serious problem in the present approach for the pole search, as the concerned resonance  poles can be easily located on the trajectory starting from the imaginary axis  by  the change of  the potential depth.

 In the following we show  for the pole spectrum of the rectangular potential how it changes with the potential depth.   The rectangular potential produces only one open trajectory  and a finite number of closed trajectories   for a finite value of $U$.  The open trajectory passes  all the attractive  resonance  poles and  almost  all of the repulsive resonance poles for the change of $\alpha$ from the  starting value on the imaginary axis  to $-\infty$   and   corresponding  conjugate complex  virtual poles to  $\infty$    in the present choice of the phase factor, while the  closed trajectory   completes with  the periodicity of $\Delta \alpha =2 \pi$ or $4 \pi$.  The open trajectory implies the existence of an infinite number of resonance-state  and  complex virtual-state poles and  its appearance is very restricted  to the case of  some ``singular" potentials as the  present rectangular potential,

\subsection{Symmetric states: poles of $S_{+}$}

% =========== Fig1 ============= %
\begin{figure}[t]
\begin{center}
\includegraphics[width=6.5cm]{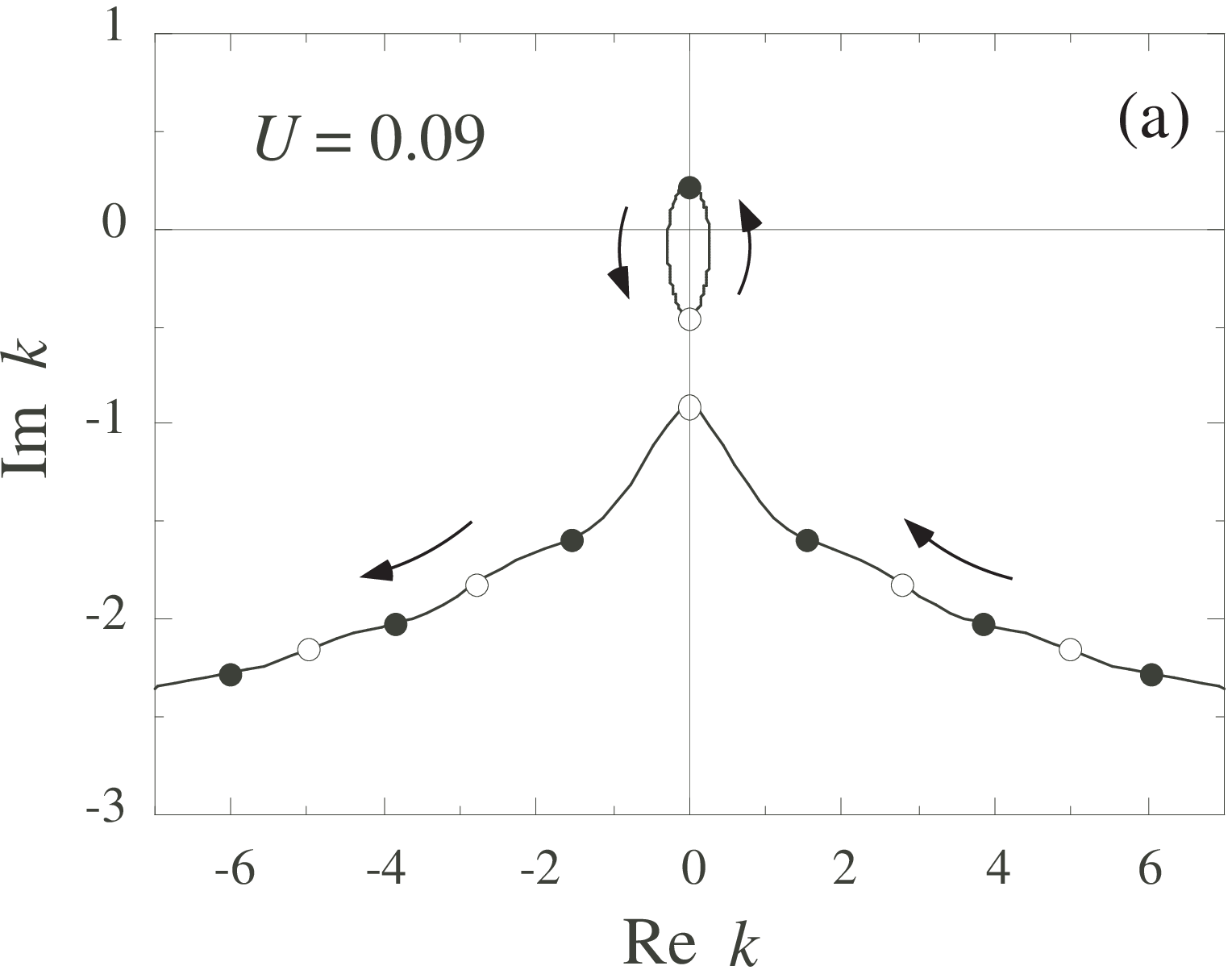}\qquad\includegraphics[width=6.5cm]
{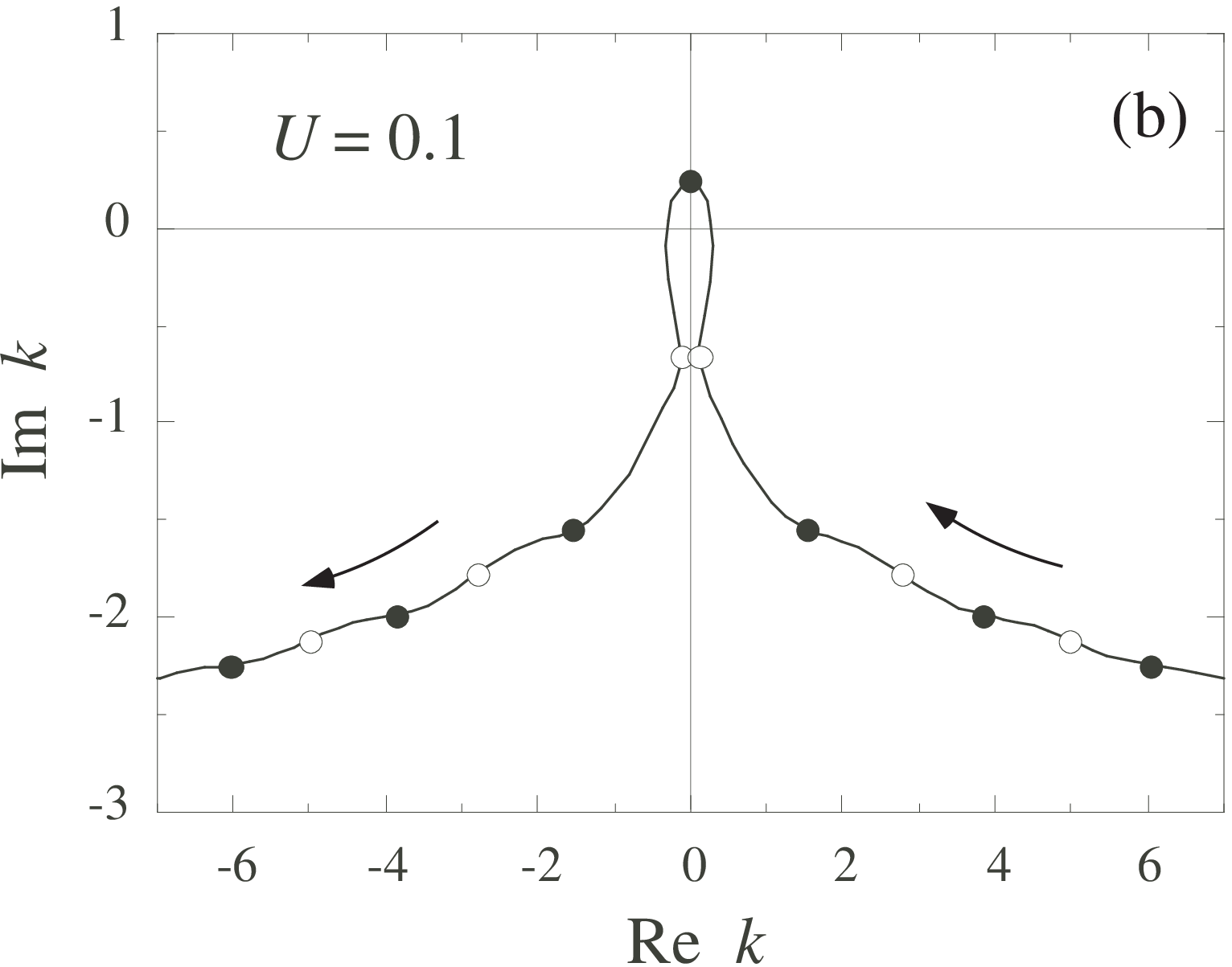}

\vspace{4ex}
\includegraphics[width=6.5cm]{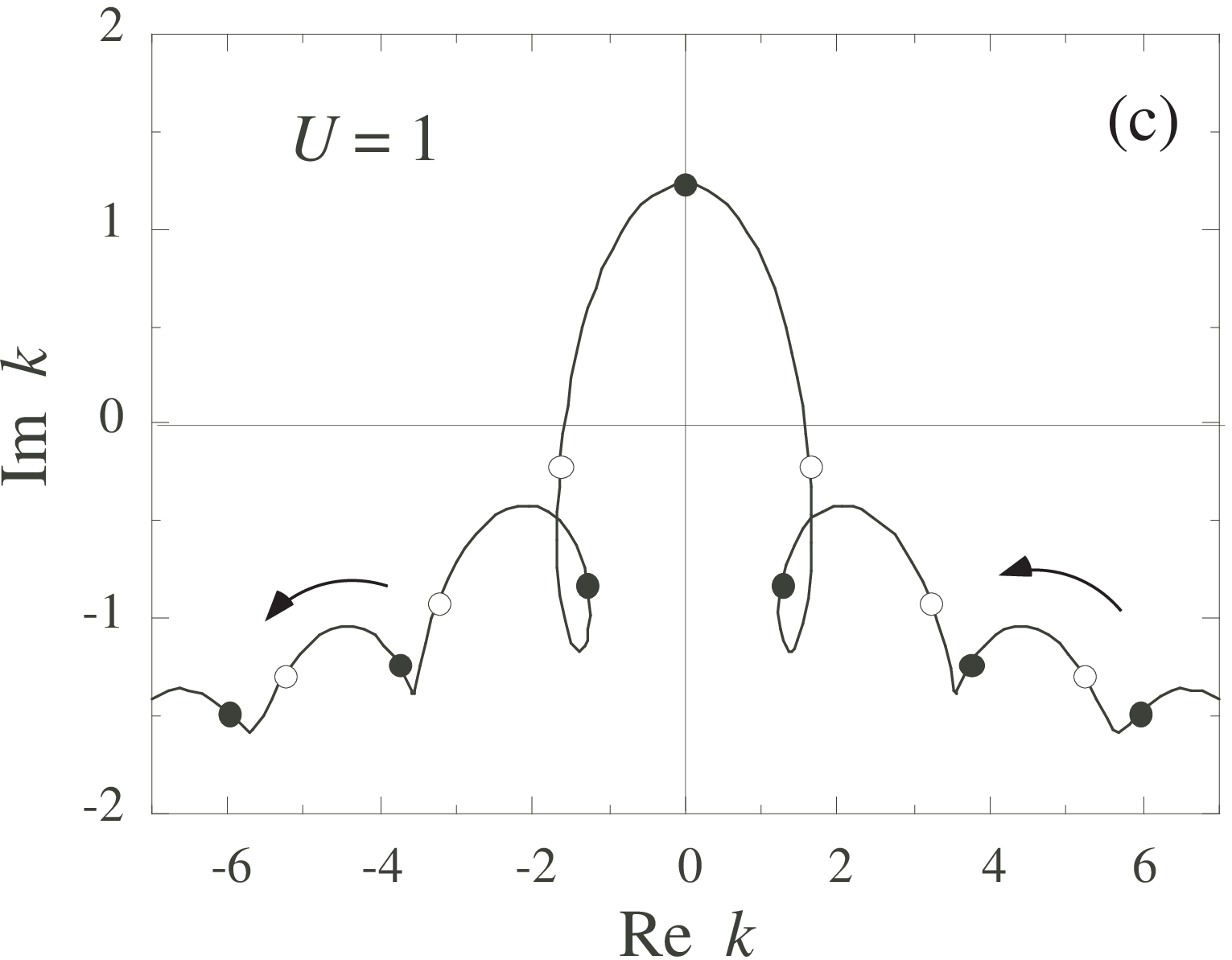}\qquad\includegraphics[width=6.5cm]
{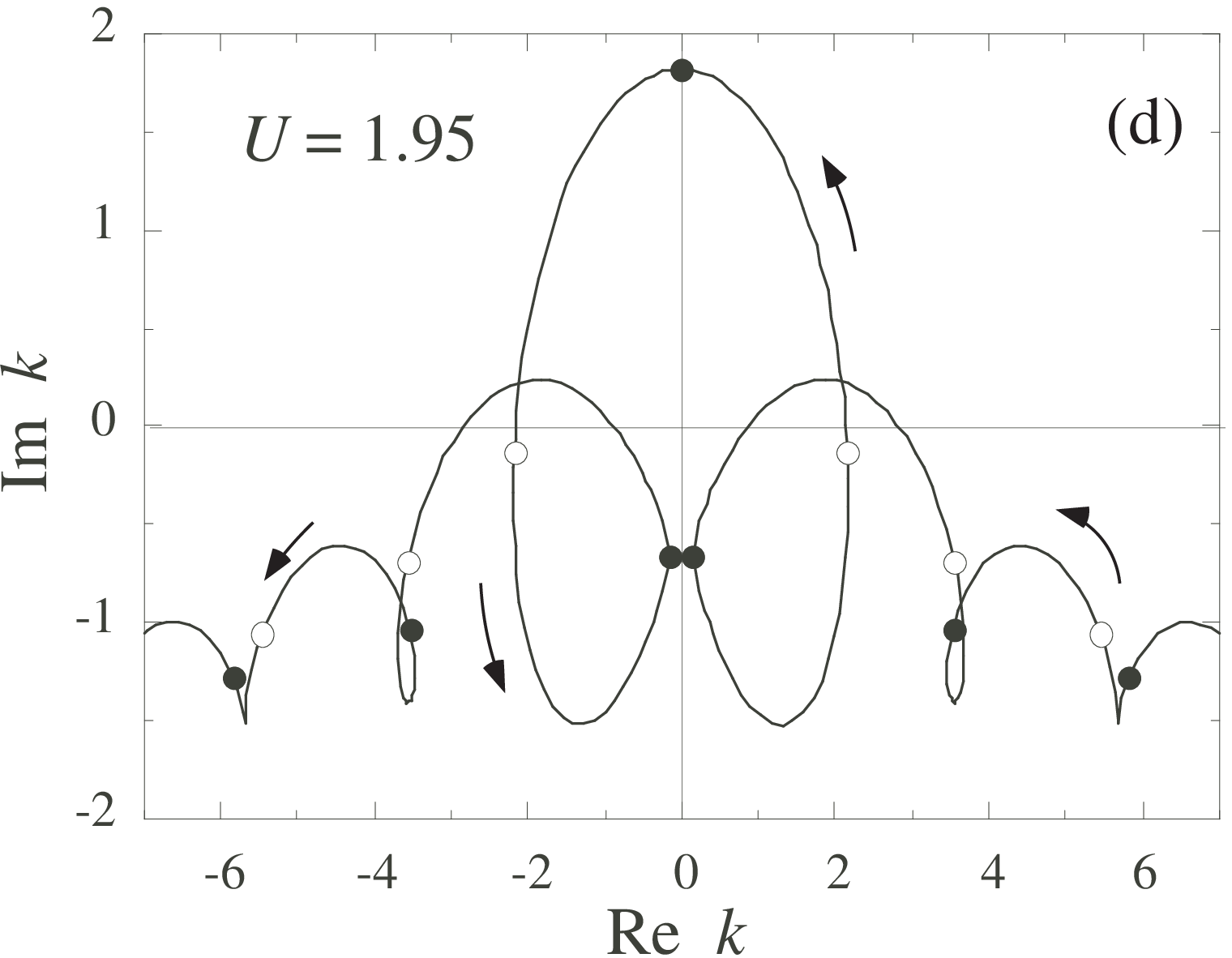}

\vspace{4ex}
\includegraphics[width=6.5cm]{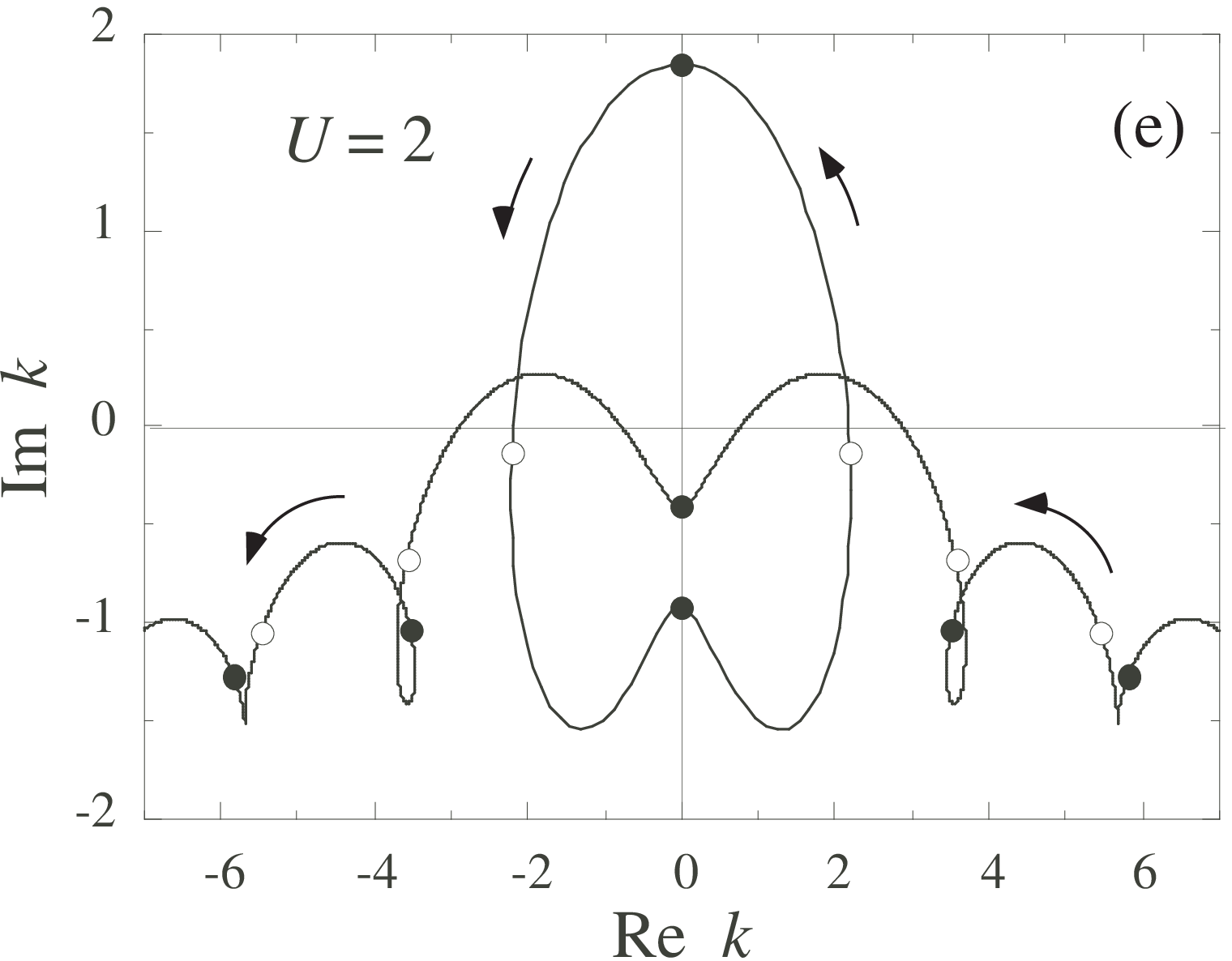}\qquad\includegraphics[width=6.5cm]{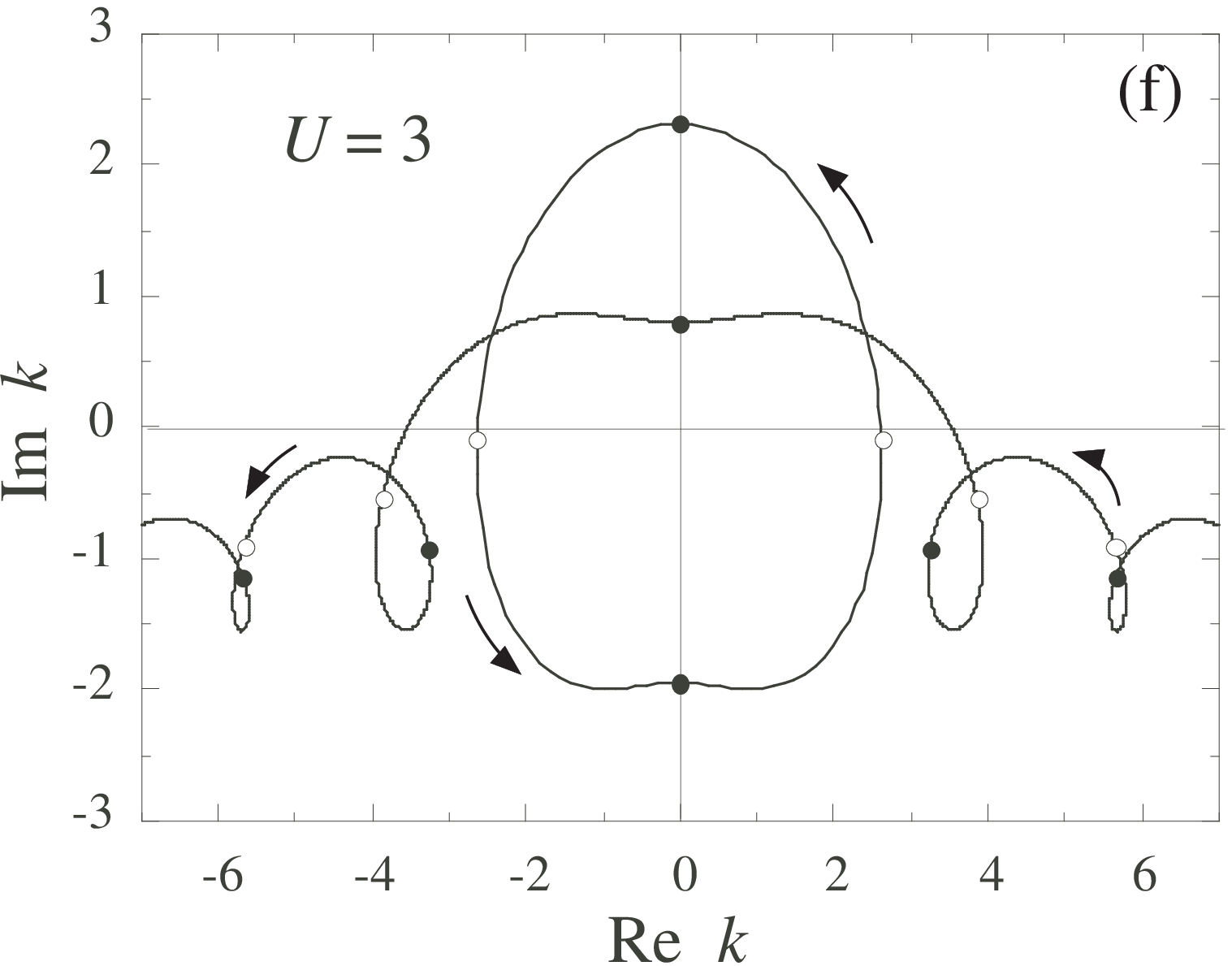}

\caption{The trajectories of the poles of  $S_{+}$ for the  symmetric states of the one-dimensional  symmetric rectangular potential. The full circles (\fullcircle)  indicate the attractive poles and the open circles (\opencircle) the repulsive poles.  The arrows given along the curves indicate the directions of the movement of poles as  the phase parameter $\alpha$ increases.
\label{fig: RECT1}}
\end{center}
\end{figure}
% ================================= %
Here we   take $m=1$ and $a=1.5$  in  the atomic units. 
The pole charts of the symmetric states  of   $S$-matrix element $S_{+}$  are given  in figure 1, where the full circles indicate the attractive poles and the open circles the repulsive poles.  The lines connecting the circles are the paths of poles  drawn by continuously changing  the phase $\alpha$. The  relation  (\ref{eq: rel3}) implies that the   complex virtual-state   poles  emerge  always in pair with  the resonance-state poles at the symmetrical  points about the imaginary momentum axis.  Any trajectory is  symmetric about the imaginary momentum axis, if it crosses the imaginary axis. If it does not cross the axis, it has a mirror symmetric partner.
 
For small $U$, there appear   one bound-state and two repulsive virtual-state poles on the imaginary  momentum axis (figure 1(a) $U=0.09$).  Starting  from these poles,  two  trajectories are given;  one is the closed  $2\pi$-periodic  trajectory which has  the bound state and one of the two repulsive virtual states  and the other is the open trajectory  having all of  the  resonance and  complex virtual states  as well as  the other repulsive virtual state. 

Necessity of {\it repulsive} virtual states for the present approach lies in the fact that some trajectories cross the imaginary momentum axis at the {\it  repulsive} virtual states; without these repulsive virtual states it is not possible to complete the  pole chart  for the corresponding attractive potential by the present procedure starting from the  poles on the imaginary momentum axis. 

As $U$ increases, two repulsive virtual states approach; the upper one going down and the lower  going up.   These repulsive virtual states finally collide  at  the momentum $k_\mathrm{c}=-\rmi/a\approx -0.667\,\rmi$ and change into a pair of a repulsive  resonance and its conjugate complex virtual state, causing  the fusion of  the closed trajectory  and the  open trajectory  into  one open trajectory (figure 1(b)  $U=0.1$).    

If $U$ increases further, poles generally go upwards, with the attractive resonance  states approaching and with  the repulsive ones leaving the imaginary axis (figure 1(c) $U=1$).  Then  a pair of attractive resonance-state  and complex virtual-state poles approach the imaginary momentum axis (figure 1(d) $U=1.95$).  For deepening the well, this pair of poles  collide  at $k_\mathrm{c}$ on the imaginary momentum axis, then turn into two attractive virtual states with a formation of one closed  $4\pi$-periodic trajectory  from the open trajectory (figure 1(e) $U=2$).  As $U$ increases,  the attractive virtual-state pole on the open trajectory  becomes an excited  bound-state pole (figure 1(f) $U=3$).

\subsection{Antisymmetric states: poles of $S_{-}$}
% =========== Fig2 ============= %
\begin{figure}[t]
\begin{center}
\includegraphics[width=6.5cm]{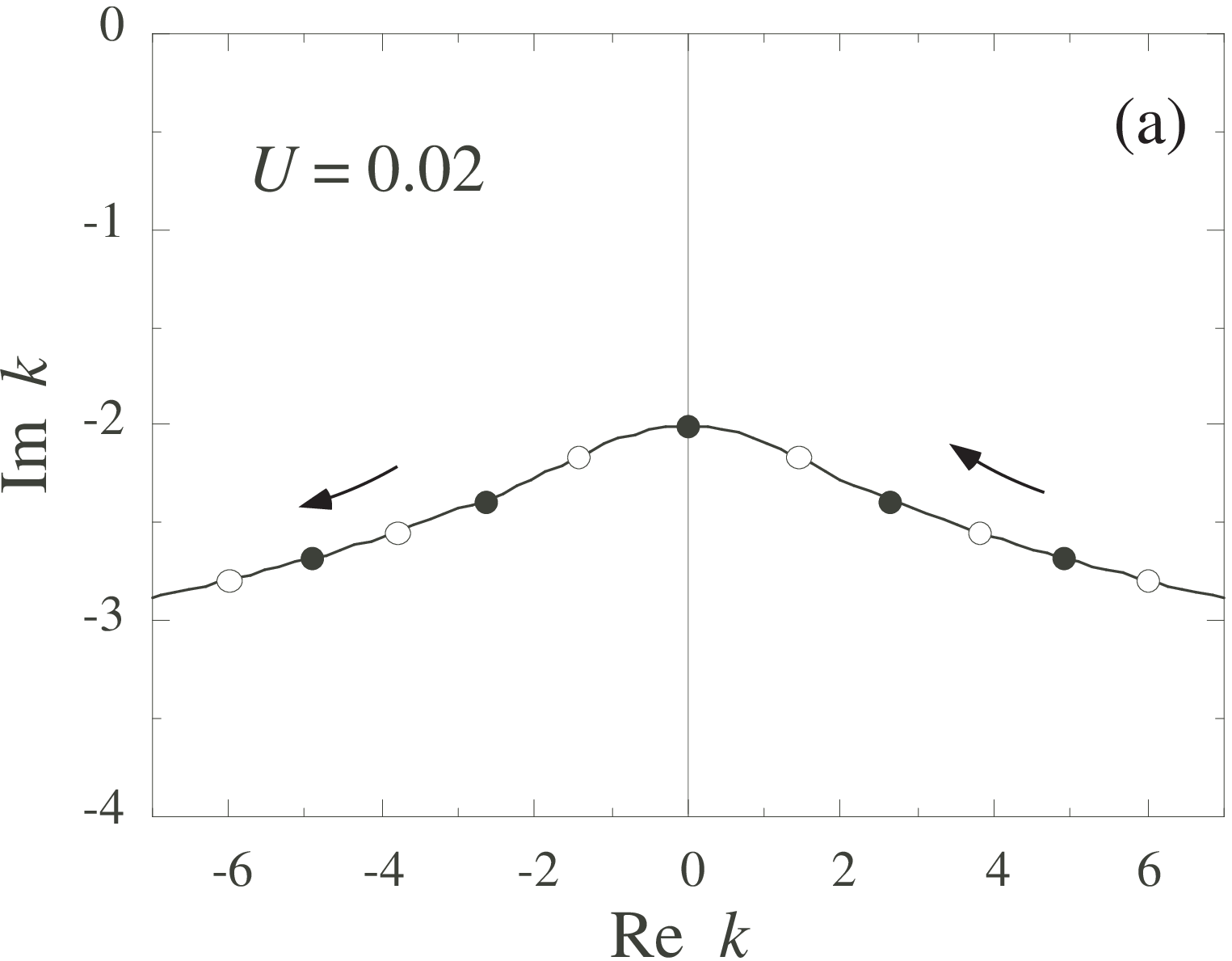}\qquad\includegraphics[width=6.5cm]
{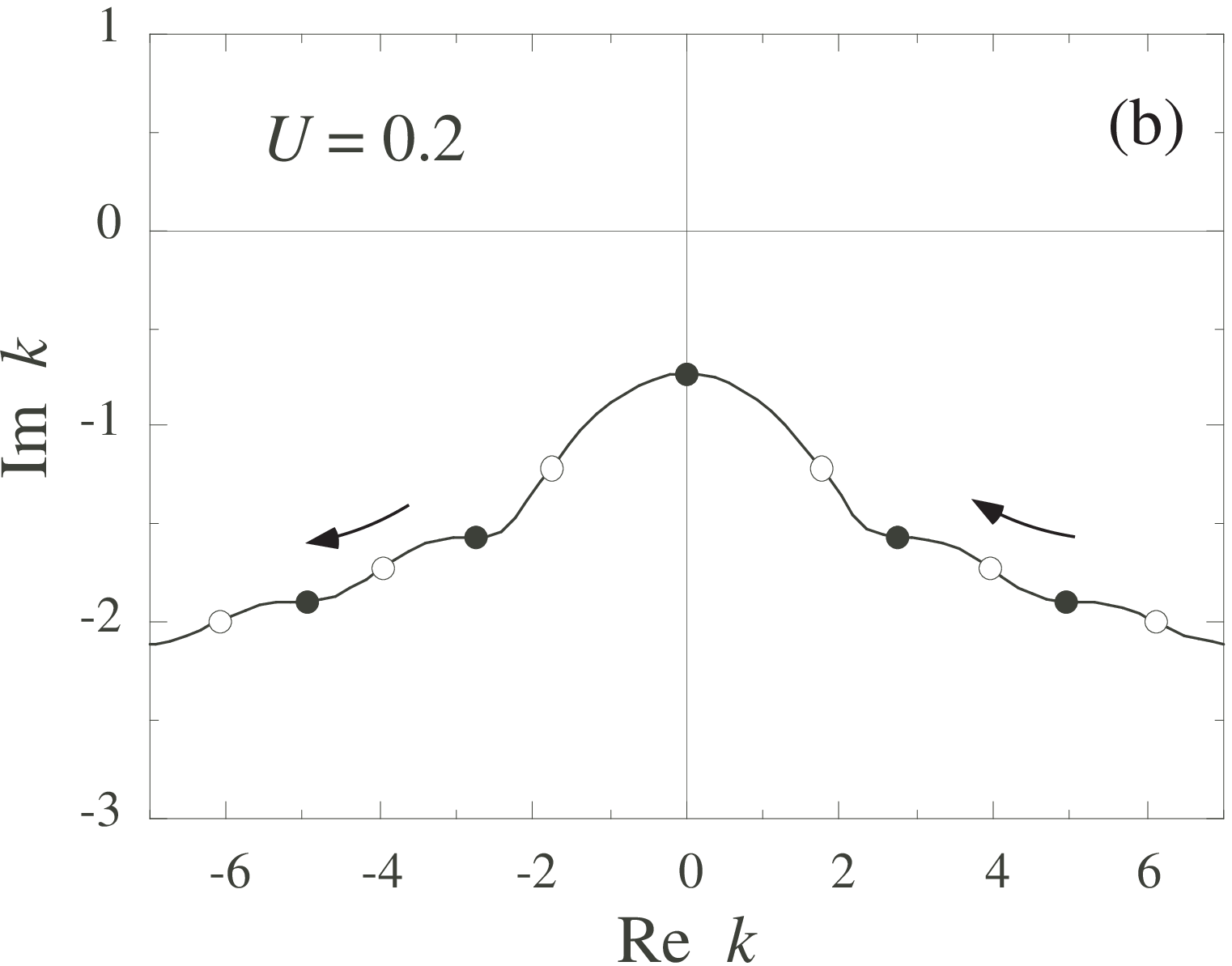}

\vspace{4ex}
\includegraphics[width=6.5cm]{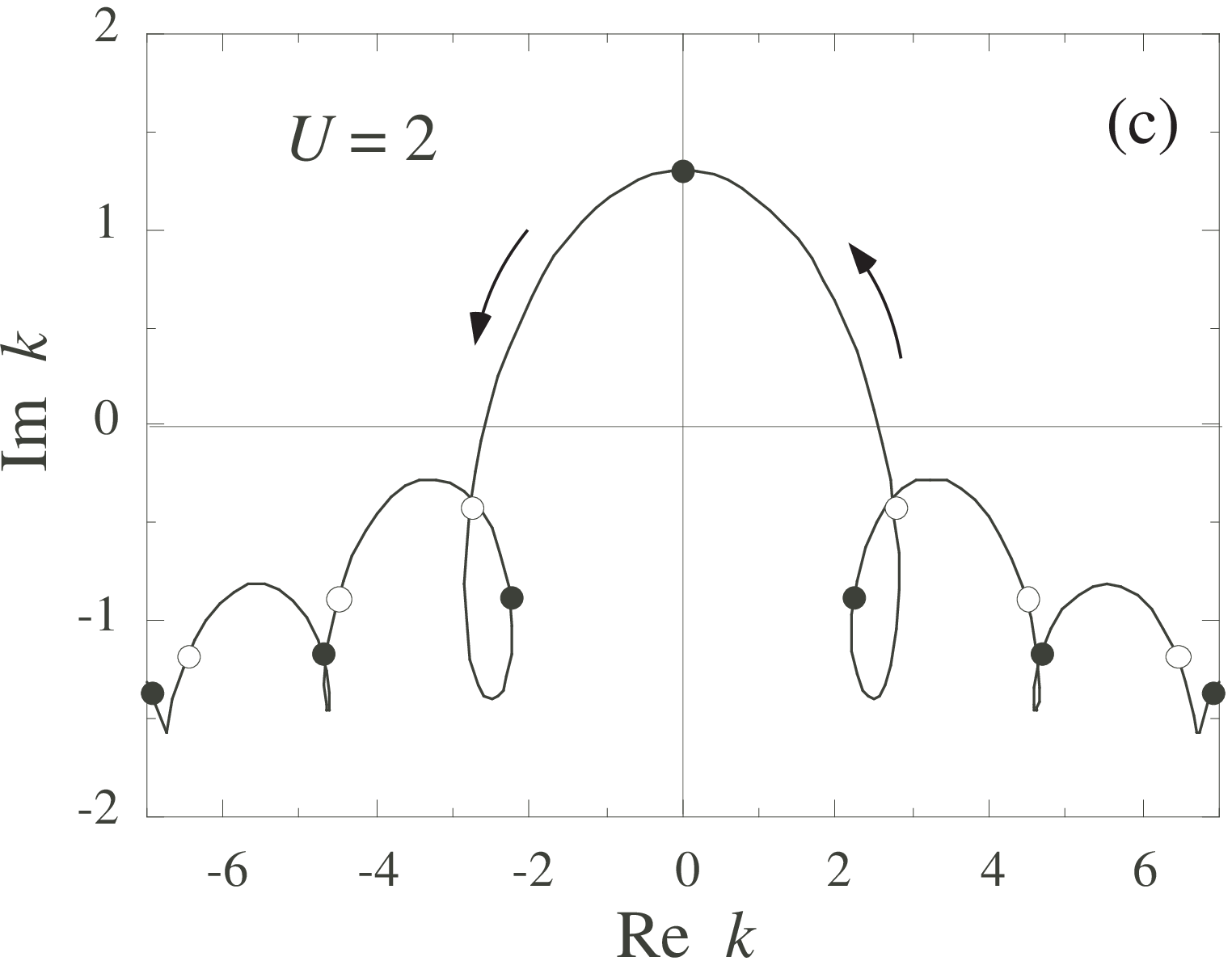}\qquad\includegraphics[width=6.5cm]
{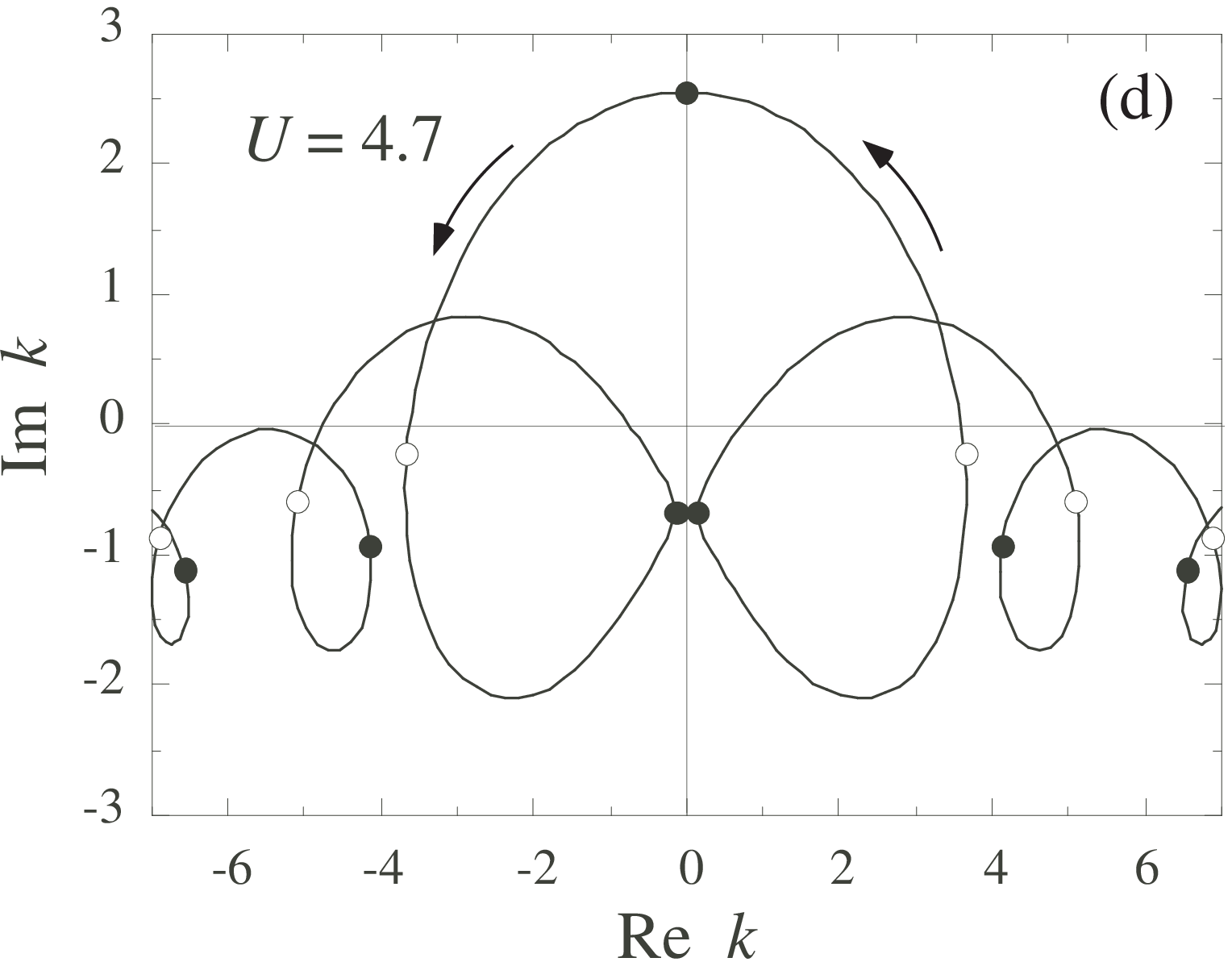}

\vspace{4ex}
\includegraphics[width=6.5cm]{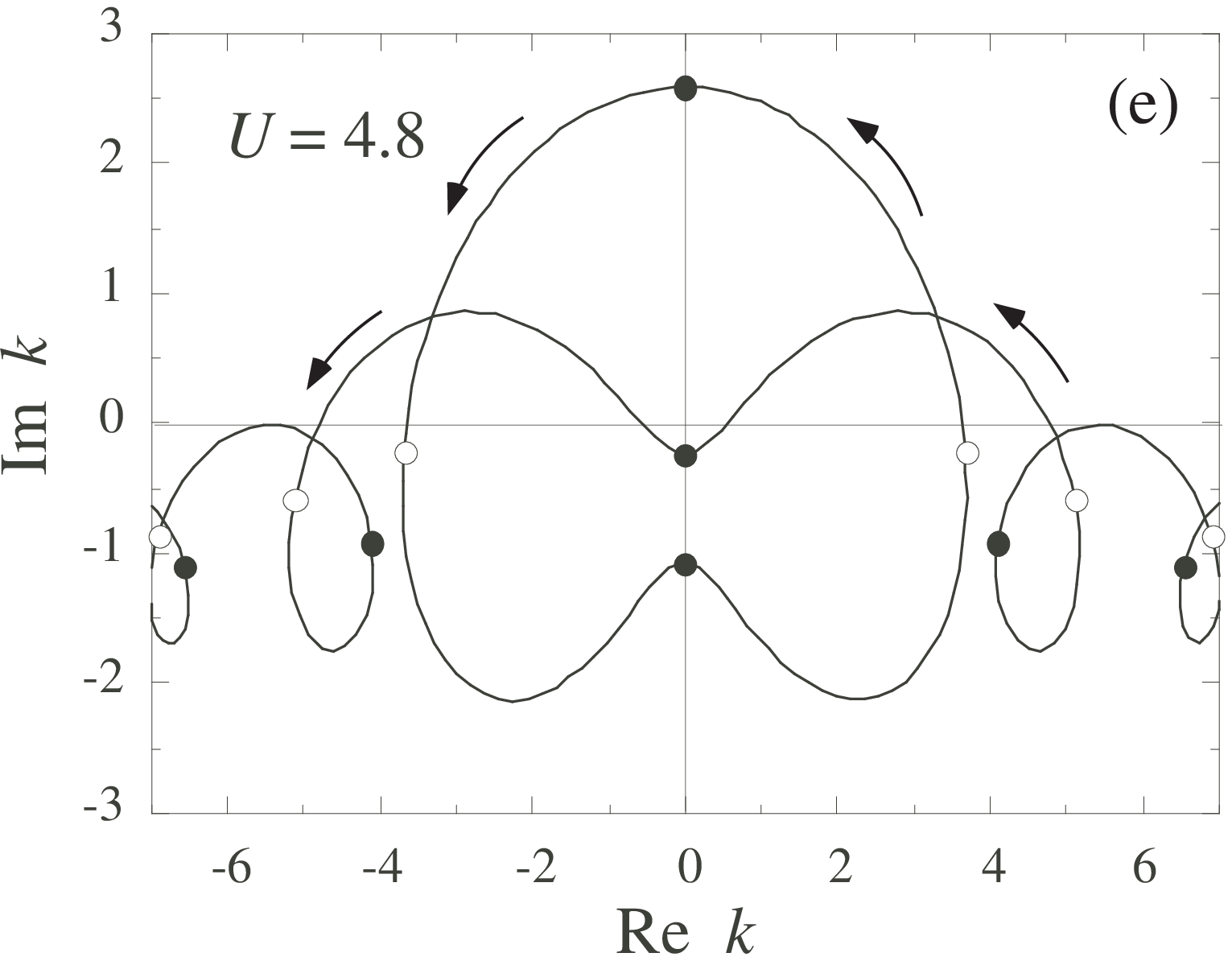}\qquad\includegraphics[width=6.5cm]{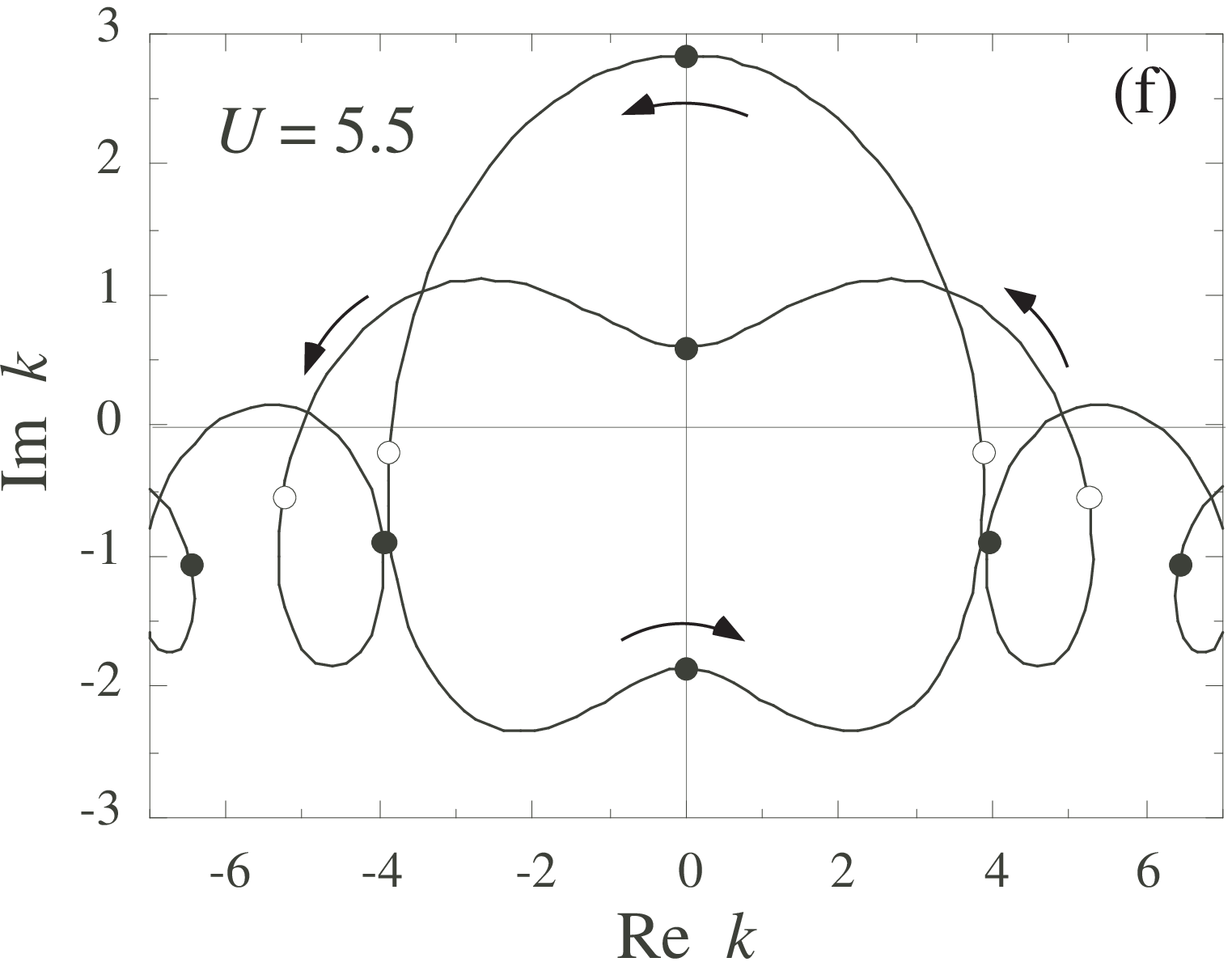}

\caption{The trajectories of the poles of  $S_{-}$ for the  antisymmetric states of the one-dimensional symmetric rectangular potential. The full circles (\fullcircle) indicate the attractive poles and the open circles (\opencircle) the repulsive poles.  The arrows indicate the directions of the movement of poles as   the phase parameter $\alpha$ increases.
\label{fig: RECT2}}
\end{center}
\end{figure}
% ================================= %
The pole charts of the antisymmetric states  of   $S$-matrix element $S_{-}$  are given  in figure 2.

For the antisymmetric states, we have only one attractive virtual-state  pole for small $U$, and there is only one open trajectory which passes all the resonance-state  and complex virtual-state poles (figure 2(a) $U=0.02$). As $U$ increases, all the poles move upwards (figure 2(b) $U =0.2 $)  with the attractive resonance-state and complex virtual-state   poles approaching  and  the repulsive ones  leaving the imaginary axis as in the symmetric states.   If $U$ becomes larger than $U_{0}^{-}=\pi ^{2}/8ma^{2}\approx 0.548$,  the lowest bound state of the antisymmetric state appears as the first excited state of the rectangular potential.  Even in this case there is only one open trajectory (figure 2(c) $U=2$)).  

As $U$ increases further, a pair of attractive resonance-state and  complex virtual-state poles  mutually approach  (figure 2(d) $U=4.7$)  and  collide on the imaginary momentum axis. Then,  these poles  turn into two attractive virtual-state poles, one going up and the other going down, forming  one $4\pi$-periodic closed and one open trajectories (figure 2(e)  $U=4.8$).  As the potential  becomes deeper,  the  attractive virtual-state pole  of the open trajectory becomes  a bound-state pole (figure 2(f) $U =5$).   As noted before,  the  collision of attractive resonance pole with its conjugate  complex virtual  pole occurs at  the point   $ k_\mathrm{c}$ which is  the common colliding spot for all the pairs of resonance-state  and complex virtual-state  poles.

The movements of  the poles of $S_{-}$  are very similar  to those of $S_{+}$ except the change of  the two repulsive virtual-state poles into a pair of  resonance-state and complex virtual-state poles in the latter. 

It is  well-known that the $s$-wave pole  spectrum  of the quantum system with the spherical rectangular  potential  with range $a$ in three-dimensional space is completely the same as  that of the antisymmetric  states of the  system with one-dimensional one.

\subsection{Applications to other potentials}

 The present procedures can be applied to any of potentials if the analytical expressions  of the $S$-matrix elements  are given explicitly. 
For example,  let us take  the exponential  \cite{exppt} and the Hulth\'{e}n \cite{htpt} potential in three-dimensional space.  These two potentials show very similar behaviours to each other for the attractive case  with an infinite number of virtual state poles for  shallow potential depth.  As the potential depth increases, these virtual-state poles go upwards on the imaginary momentum axis and   successively emerge on the positive imaginary momentum axis as bound-state poles. The spectra for the repulsive case, however,  are different. The Huth\'{e}n potential gives an infinite number of virtual-state poles which are simply going downwards as  the potential strength increases, while the  repulsive virtual states for the exponential  turn successively into resonance-state and complex virtual-state  poles.  All  the trajectories  of the Hulth\'{e}n  potential are closed  $2\pi$-periodic  ones, while the exponential potential gives  a finite number of $4\pi$-periodic  trajectories  and an infinite  number of  $2\pi$-periodic   ones. These can be  easily observed  by the present approach. Some details of the  pole trajectories for the three-dimensional spherical symmetric potentials including  the exponential and Hult\'{e}n  potentials will be given elsewhere.

\section{Some Remarks}

 The  bound states, the virtual states and the resonance states including the conjugate complex  virtual states have  the same dynamical origin and  are mutually transforming as the potential strength  varies.  

 In practice,  with  currently available accessibility  to computing facilities,   the resonance-pole research  does not require  laborious work even in the approach by Nussenzveig \cite{nssn}, as far as the analytical expressions of the $S$-matrix are given as in the case of the rectangular potential.  The method  of complex extension of potential  presented in this article  affords a simple and systematic way of making  a global pole chart for a given potential and  of observing the change of spectrum as the potential strength varies.

We  have shown the trajectory structures of the pole  spectrum for the attractive rectangular potential.  Here the  \textit{repulsive} resonance-state and  complex virtual-state poles play an important role as  mediators connecting attractive  poles.  Even without considering such role of the repulsive resonance states, it seems necessary to pay more attention to these states, especially of the $s$-wave ones, which seem to appear commonly  for the most of repulsive potentials,  as observed for a group of  potentials having the exponential tail.  These poles could appear very near the real energy axis at low energies. 

The present approach relates  two physical systems having the potentials  of the same shape but with opposite signatures  and  will be useful to consider the quantum system from an extended global aspect. The physical and mathematical implications of  the pole-trajectory structures  need further investigations.

\section*{References}
%**************************** %

%**************************** %

% ************************************ %

\begin{thebibliography}{99}
%**************************** %
\bibitem{zvms} Zavin R and Moiseyev N 2004 {\it J. Phys. A: Math. Gen.}
{\bf 37} 4619--28
\bibitem{nssn} Nussenzveig H M 1956 {\it Nucl. Phys.} \textbf{11} 499--521
\bibitem{cxsmrl} Joffily S 1973 {\it Nucl. Phys.} {\bf A215} 301
\nonum Kok L P and Van Haeringen  H 1981 \APNY {\bf 131} 426

\bibitem{cmxcd} Aguilar J and Combes J M  1971 {\it Commun.  Math. Phys.} \textbf{22} 269
\nonum Balslev E and Combes J  M  1971{\it  Commun.  Math. Phys.} \textbf{22} 280
\nonum Moiseyev N 1998 {\it Phys. Rep.} \textbf{302} 211--93

\bibitem{riss} Riss U Vand Meyer H D 1993 {\it J. Phys. B: At. Mol. Opt.
Phys.} {\bf 26} 4503--35
\nonum Lefebvre R, Sindelka M and Moiseyev N 2005 {\it Phys. Rev.} A
{\bf 72} 052704
\bibitem{kkln} Kukulin V I, Krasnopol'sky V M and Hor\'{a}\v{c}ek J 1989
{\it Theory of Resonances; Principles and Applications} (Dordrecht
Netherlands: Kluwer Academic Pub) Chap. 5
\bibitem{nwtn} Newton R G 2002 {\it Scattering Theory of Waves and
Particles} (New York: Dover)
\bibitem{qmb} Schiff L  I 1968 {\it Quantum Mechanics} 3rd edn (New York:
McGraw-Hill)

\bibitem{exppt} Sitenko A G 1991 {\it Scattering Theory} (Springer
Series in Nuclear and Particle Physics) ed M K Gaillard {\it et al}
(Berlin$\cdot$Heidelberg: Springer-Verlag)
\bibitem{htpt} Hulth\'{e}n L 1942 {\it Ark. Mat. Astron. Fys.} {\bf 28A} 5
\nonum Hulth\'{e}n L 1942 {\it Ark. Mat. Astron. Fys.} {\bf 29B} 1
%**************************** %
\end{thebibliography}
\end{document}